# One-shot phase image distinction of plasmonic and dielectric nanoparticles

Lisa Saemisch[1], Niek F. van Hulst[1,2] and Matz Liebel[1,*]

*[1] ICFO-Institut de Ciencies Fotoniques, The Barcelona Institute of Science and Technology, 08860 Castelldefels, Barcelona, Spain*

*[2] ICREA-Institució Catalana de Recerca i Estudis Avançats, 08010 Barcelona, Spain*

\* Corresponding author: Matz Liebel;  e-mail: matz.liebel@icfo.eu

**Abstract: Nanoscale phase-control is one of the most powerful approaches to specifically tailor electrical fields in modern nanophotonics. Especially the precise sub-wavelength assembly of many individual nano-building-blocks has given rise to exciting new materials as diverse as metamaterials, for miniaturizing optics, or 3D assembled plasmonic structures for biosensing applications. Despite its fundamental importance, the phase-response of individual nanostructures is experimentally extremely challenging to visualize. Here, we address this shortcoming and measure the quantitative scattering phase of different nanomaterials such as gold nanorods and spheres as well as dielectric nanoparticles. Beyond reporting spectrally resolved responses, with phase-changes close to π when passing the particles' plasmon resonance, we devise a simple method for distinguishing different plasmonic and dielectric particles purely based on their phase behavior. Finally, we integrate this novel approach in a single-shot two-color scheme, capable of directly identifying different types of nanoparticles on one sample, from a single widefield image.**

Phase, an elusive quantity in day-to-day life, is one of the most crucial parameters in physics and the life-sciences. Historically, measurements based on phase-contrast have proven to be extremely powerful and highly sensitive to tiny changes in the morphology of structures[1] even if the underlying absolute phase-value remained unobserved. Since these pioneering works, technology has advanced rapidly and different phase-detecting techniques are of crucial importance across many scientific fields. In nanoscale physics, the phase of a nanoparticle (NP) accesses its nanoscale response[2–5] and is a crucial parameter for tailoring complex photonic structures such as metasurfaces which manipulate the amplitude, direction and polarization of light at nano- to micro-scales[6,7]. In the biomedical sciences, phase-based measurements are often key enablers for ultrasensitive measurements[8] such as protein-binding to plasmonic structures[9–11] and the emerging field of quantitative phase imaging is expected to dramatically contribute to future diagnostics[12,13].

However, even though macro-scale phase measurements are relatively easy to implement, determining the absolute scattering phase of individual nano-objects is far from trivial as one has to consider the phase of the observation wave and account for the sub-diffraction limited nature of the particle. A versatile experimental approach to directly and reproducibly determine the absolute phase of such objects would both advance the observation-based development of nanophotonic materials and allow implementing novel imaging methodologies that exploit absolute phase measurements as a robust contrast mechanism for distinguishing and identifying different materials in a complex environment.

Here, we implement such a platform in the form of a widefield off-axis holographic microscope that detects the absolute scattering phase of many individual NPs simultaneously, in a wavelength resolved fashion, which allows distinguishing between plasmonic and dielectric particles. Furthermore, we extend our approach to a single-shot-technique, which rapidly identifies many different NPs from a single widefield phase-image.

**Experimental Setup**

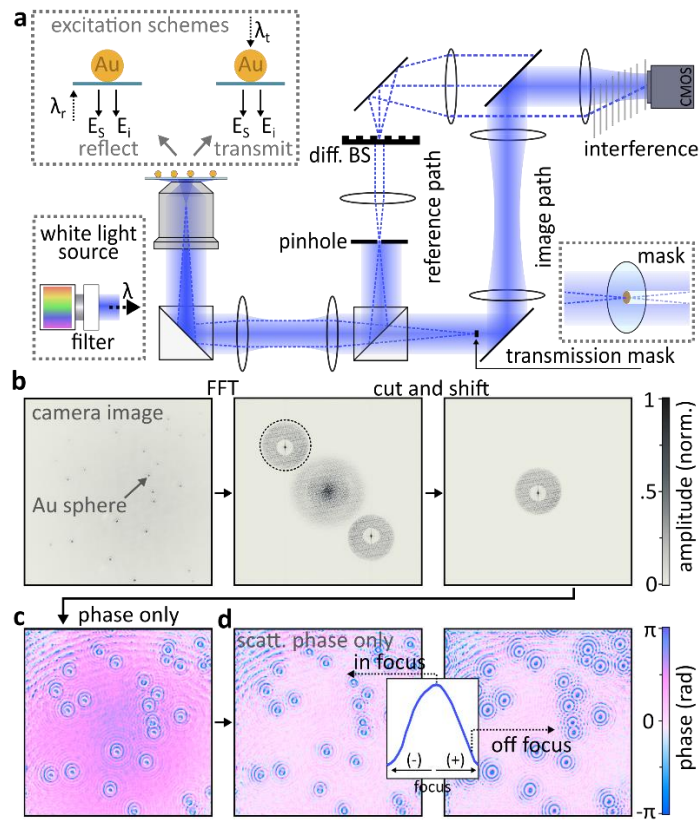

**Figure 1, Widefield quantitative phase imaging**. a) Schematic of the widefield off-axis interferometric microscope. Two possible excitation schemes are shown: reflection (illumination: $\lambda_R$) and transmission (illumination: $\lambda_T$), the scattered field is indicated as $E_s$ and the partially reflected, or transmitted, illumination wave as $E_i$. A spectrally filtered white light source illuminates the NPs and both the illumination wave (blue dotted line) as well as the NP-scattering (blue blurred line) are guided through a relay imaging system until being separated into a reference and image path. A pinhole in the reference path eliminates the image information to create the reference wave, which passes a diffractive beamsplitter and its first diffraction order is interfered on the camera with the image wave. In the image path the illumination wave is selectively attenuated with a transmission mask (inset right lower edge), consisting of a small disk of gold evaporated onto a glass window. b) Analysis procedure: the raw image (left) is Fourier transformed (centre) to isolate the interference term (dashed circle) which is then shifted to zero momentum to remove the high-frequency modulation (right). c) Isolated raw phase with residual background phase (left). The image system imperfections are removed by fitting the residual phase background (centre) and we further propagate the image digitally to find the optimal focus, which we define as the position of largest NP scattering amplitude.

Figure 1a schematically depicts the experimental setup which enables wavelength resolved phase measurements by tuning the spectrally filtered output of a supercontinuum laser across the desired frequency range. The laser widefield illuminates the sample of interest, either by passing through the objective, in a so-called reflection geometry, or by free-space illumination, in a transmission geometry (Figure 1a, inset). An oil-immersion objective then collects sample scattering as well as the, reflected or transmitted, illumination wave.

The collected light is relay imaged and divided by a 50:50 beam splitter to generate the so-called reference- and image-paths. The latter is composed of a three-lens imaging system which, ultimately, generates a real-space image of the sample on the camera. To increase the contrast between the, often weak, sample scattering and the residual illumination wave we, furthermore, insert a semi-transmissive Fourier-filter into a plane conjugate with the back-focal-plane (BFP). This filter selectively attenuates the illumination wave's amplitude by approximately 30-fold (Figure 1a, Methods)[14]. The reference-path uses an identical three-lens system but with a pinhole placed into the conjugate BFP[15]. This arrangement essentially deletes all sample-related information, thus generating a plane-wave at the position of the camera. A diffractive beamsplitter (grating) is then used to recombine reference- and image-paths, as shown in Figure 1a, and the two waves interfere on the camera. This arrangement can be regarded as an off-axis interferometer which in-situ generates its reference wave[15] and, furthermore, allows independent image manipulation by additional Fourier filtering. As a result, we are able to obtain NP phase-responses and relate the measured quantities to a simultaneously measured reference value, obtained at sample locations which do not contain NPs. Ultimately, this arrangement allows us to extract the absolute particle scattering phase.

**Image formation and interference**

To simplify the discussion we will, initially, ignore interference between the image- and the reference-path, which allows us to describe their respective camera images $I_{img}(x,y)$ and $I_{ref}(x,y)$ as:

$$I_{img}(x,y) = A_s^2(x,y) + \frac{A_i^2(x,y)}{a^2} + \frac{2A_sA_i}{a}\cos(\Delta\varphi)(x,y) \tag{1}$$

and

$$I_{ref}(x,y) = A_{ref}^2(x,y) \tag{2}$$

, with $x$ and $y$ describing the position, $A_s$ and $A_i$ the electric field amplitudes of the scattering and illumination waves, $\Delta\varphi$ the phase-difference between the two waves and $a$ the attenuation amplitude of the semi-transparent transmission mask. $A_{ref}$ is the amplitude of the reference wave, which is generated by passing the illumination wave through the pinhole in the BFP as described above (Figure 1a). We experimentally select the illumination wave attenuation to ensure $A_s \gg \frac{A_i}{a}$, which further simplifies the mathematical description of the $I_{img}$ to:

$$I_{img}(x,y) \approx A_s^2(x_p, y_p) \quad \text{(if a NP is present at } x,y)$$
$$= \frac{A_i^2(x_{nop}, y_{nop})}{a^2} \quad \text{(if no NP is present at } x,y) \tag{3}$$

, where $(x_p, y_p)$ denote image-regions where scattering particles are located and $(x_{nop}, y_{nop})$ regions without particles. Following this simplification, we are able to describe the holographically recorded image, considering $A_s$ and $A_i$ and $A_{ref}$, as:

$$I_{hol}(x,y) = A_s^2(x_p, y_p) + \frac{A_i^2(x_{nop}, y_{nop})}{a^2} + A_{ref}^2(x,y)$$
$$+ 2A_{ref}\,A_s\cos(\Delta\varphi_{sref})(x_p, y_p) + \frac{2A_{ref}A_i}{a}\cos(\Delta\varphi_{iref})(x_{nop}, y_{nop}) \tag{4}$$

, where the phase-difference terms can be described as:

$$\Delta\varphi_{sref} = \left(\varphi_s(x_p, y_p) - \varphi_{ref}(x_p, y_p) - \varphi_{gx} * x - \varphi_{gy} * y\right) \tag{5}$$

and

$$\Delta\varphi_{iref} = \left(\varphi_i(x_{nop}, y_{nop}) - \varphi_{ref}(x_{nop}, y_{nop}) - \varphi_{gx} * x - \varphi_{gy} * y\right) \tag{6}$$

, with $\varphi_s(x_p, y_p)$, $\varphi_i(x_{nop}, y_{nop})$ and $\varphi_{ref}(x,y)$ being the position dependent phase of the scattered, illumination and reference wave, respectively. The camera-chip position dependent terms $\varphi_{gx} * x$ and $\varphi_{gy} * y$ are grating induced, linear, phase-ramps that form the basis for off-axis holography based interference term isolation[16].

We will now outline three additional steps that are necessary to isolate the pure scattering phase from the experimentally acquired hologram (Equation 4): i) isolation of the interference terms; ii) removal of the reference and illumination phase and iii) accounting for de-focus and setup-induced phase effects.

**Interference term isolation**

Figure 1b conceptually outlines the steps taken to isolate the interference term of interest. A fast Fourier transformation of the experimentally obtained image yields its momentum-space representation (Figure 1b), where the amplitude-square terms are located in the center and the interferometric terms are visible as diagonally-shifted mirror images. We isolate one of the interference terms and multiply the cut image by a unity circle to set the frequency background around the interference term to zero. Next, we shift the interference term to the center (Figure 1b), which is equivalent to removing the two linear-phase terms $\varphi_{gx}$ and $\varphi_{gy}$ introduced via the grating (Equations 5, 6). Inverse Fourier transformation then isolates the complex image, $I_{filtered}(x,y)$:

$$I_{filtered}(x,y) = A_{ref}(x,y)[A_s(x_p, y_p)e^{-i\left(\varphi_s(x_p,y_p)-\varphi_{ref}(x_p,y_p)\right)}$$
$$+ \frac{A_i}{a}(x_{nop}, y_{nop})e^{-i\left(\varphi_i(x_{nop},y_{nop})-\varphi_{ref}(x_{nop},y_{nop})\right)}] \tag{7}$$

**Removing the reference phase**

Equation 7 shows that we can directly separate the image into its amplitude and phase terms but to obtain the pure scattering phase of a particle we need to both remove the residual reference phase, $\varphi_{ref}(x,y)$, and account for the fact that the measured scattering phase is intrinsically coupled to the phase of the illumination wave:

$$\varphi_s(x_p, y_p) = \varphi_{s\_final}(x_p, y_p) + \varphi_i(x_p, y_p) \qquad (8)$$

, with $\varphi_{s\_final}(x_p, y_p)$ being the pure scattering phase of interest. To access this information, we thus need to subtract the illumination wave's phase, $\varphi_i(x_p, y_p)$, described in the second term of Equation 7. To obtain the necessary information, we assume that the illumination wave's phase measured next to a scattering particle is equivalent to its phase at the particle's position. This is a justified assumption, since both the reference- and illumination- phase are strongly confined in momentum space and hence vary slowly. We consequently describe this slowly varying residual phase (Figure 1c) using Zernike polynomials (n=0-3: piston, tilt, defocus, astigmatism, coma and trefoil) and, following subtraction, obtain an image with a flat phase, close-to-zero, over the entire image (Figure 1d). Additional division by the, separately measured, reference amplitude $A_{ref}$ then yields the final image:

$$A(x, y) = A_s(x_p, y_p)e^{-i\left(\varphi_{s\_final}(x_p, y_p)\right)} + \frac{A_i}{a}(x_{nop}, y_{nop}) \qquad (9)$$

**Focusing and attenuation mask**

Lastly, we computationally refocus the image using the angular spectrum method[17], where we define the focus as the z-position of maximum particle-amplitude (Figure 1d, inset). Furthermore, we apply an additional phase-offset to account for the semi transmissive mask, acting as Fourier-filter. Said filter adds additional material, in the form of a *d=120 nm* gold film, to the optical path of the illumination wave. Since only the illumination wave passes the thin gold film, while the scattering wave propagates through air, we simply calculate the phase-difference via:

$$\varphi_{diff}(\lambda) = 2\pi(n_{gold}(\lambda) - n_{air}(\lambda))\frac{d}{\lambda} \qquad (10)$$

, where $n_{gold}$ and $n_{air}$ are the respective refractive indices, *d* the thickness of the gold film and $\lambda$ the wavelength. We note that even though a description relying on Fresnel equations could be used, we opted for the mathematically less complex version outlined above as we did not note any significant deviations for the relatively thick film used here, even for different film-materials (Supplementary Information). After accounting for the mask thickness, we extract the absolute scattering-phase of the individual NPs as the phase value in the center of each particles' point-spread-function.

**Experimental benchmark on nanorod antennas**

We benchmark the performance of our experiment by employing arrays of gold nanorods with a width/height of 50 nm and lengths increasing from 70 nm to 220 nm, in steps of 10 nm, fabricated by e-beam lithography (Methods). These samples provide plasmonic resonances that span from the visible to the near infrared, which make them an ideal candidate for ensuring that spectrally resolved scattering phases can be obtained in an artifact-free manner.

Figure 2a compares amplitude and phase images, recorded for an array of nanoantennas, using a transmission geometry and wavelengths of 615 nm and 875 nm, respectively. The shorter wavelength is off-resonant for the nanoantenna length-range and all phases show similar values, around $-\pi/2$. Using 875 nm illumination we observe a scattering amplitude-maximum for the 150/160 nm-long nanoantennas. As we pass the resonance, we identify a phase-flip of approximately $\pi$ with shorter antennas exhibiting $\pi/2$ phases and longer antennas values around $-\pi/2$. These observations agree with the model of a driving field that acquired a $\pi/2$ Gouy-phase shift due to scattering off a NP[18]. The intuitively expected $\pi$ to zero phase-change, when passing the resonance from blue (higher frequency) to red (lower frequency), is thus shifted to $3\pi/2 - \pi/2$ which is equivalent to $-\pi/2 - \pi/2$ after phase-wrapping. Importantly, the Gouy-phase shift mentioned above is solely a near-field effect and additional Gouy-phase shifts when propagating the imaging system might occur but will not impact our measurement as both the scattered as well as the illumination waves acquire identical shifts, albeit at different planes in the relay-imaging system.

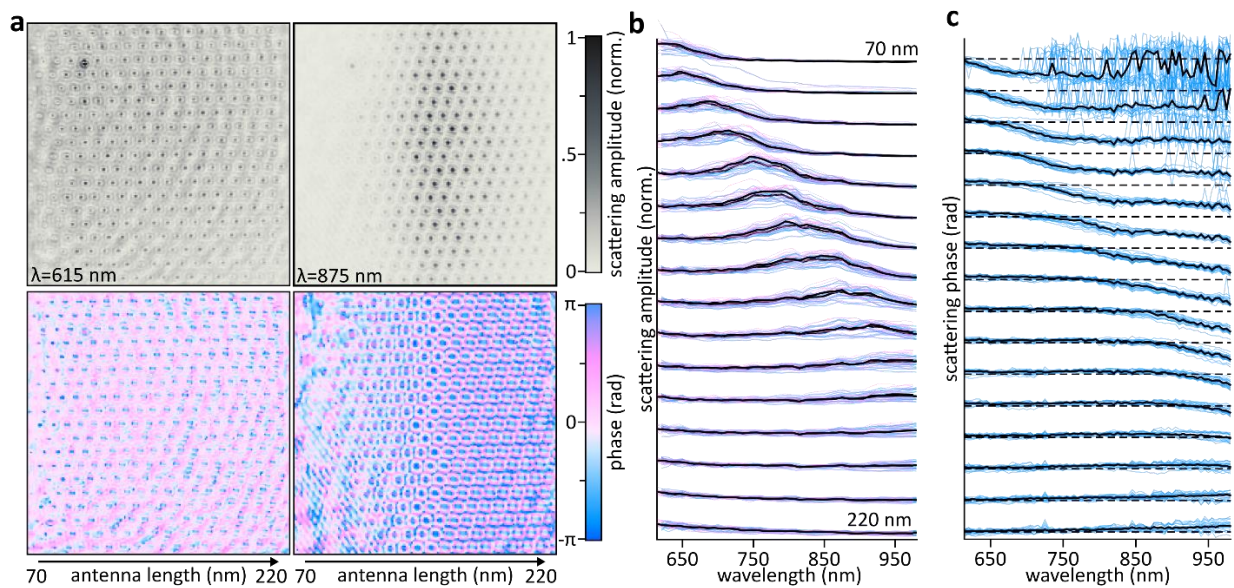

**Figure 2, Phase and amplitude properties of gold nanorods of different lengths.** a) Amplitude (top) and phase (bottom) images of gold nanoantennas with lengths increasing from 70 nm to 220 nm for two different excitation wavelengths, 615 nm (left) and 875 nm (right). b) Scattering amplitude at different illumination wavelengths for all nanorods seen in a), length increases from top to bottom. The color indicates forward scanning of the illumination wavelength (615 to 975 nm in steps of 10 nm, pink) and backward scanning (980 to 620 nm, blue). The median-values of the two scanning procedures are shown in black. c) Phase values for the different nanorods from b), combining the forward and backward scanning experiments. Single nanorods are seen in blue, the median for each length is presented in black. The individual phase curves are offset by $\pi$, the dashed line indicates $\varphi = -\pi/2$ for the respective rod-length. Hence the nanorod phase change is observed from $\varphi = -\pi/2$ to $\varphi = \pi/2$, considering the phase wrapping from $3\pi/2$ to $\pi/2$. For representation purposes, we moved the point of phase wrapping from $\pi$ to 1.5 rad to avoid phase wrapping as we move through the particles' resonance.

To gain further insight, we record the scattering amplitude and phase of all antennas as a function of illumination wavelength. To gauge the reproducibility of our measurements and phase corrections, we perform two measurements. First, we tune the illumination wavelength from 615 nm to 975 nm and then from 980 nm to 620 nm, both in steps of 10 nm in ascending and descending order, respectively. Figure 2b compares the amplitude data of both scans and highlights the expected scattering resonance red-shift as we increase the antenna-length. At the same time, we observe a phase flip of approximately π as we move across the resonance of the individual antennas (Figure 2c). We note that the dramatic phase-fluctuations observed for shorter nanoantennas in the near-infrared spectral region are due to insufficient signal-to-noise ratio, for these weakly scattering objects, and not an intrinsic limitation of our experimental approach. Importantly, the data presented in Figure 2b, c is composed of two consecutively performed experiments, with 5 nm shifted illumination wavelengths, which highlights the robustness of our phase-extraction and correction approach.

**Gold and dielectric nanoparticles**

Even though the lithographically fabricated nanorods are an ideal model system for gauging the robustness and operational reliability of our approach, most biophysical applications rely on considerably smaller spherical NPs, which are often detected in a so-called back-scattering, or reflection, configuration, which avoids propagating through the sample[19–21]. Additionally, this configuration conveniently reduces the amplitude of the residual illumination wave, as only a fraction is back-reflected into the microscope objective, thus yielding favorable higher scattering to illumination wave ratios. To mimic the experimental conditions of a live-cell experiment where scattering from Au NPs alongside scattering-background of the dielectric cellular environment are simultaneously present, we perform phase and scattering amplitude measurements of Au and polystyrene (PS) nanospheres, with respective diameters of 60 nm and 100 nm, using the reflection geometry. Said particles are immobilized by spin coating an aqueous polyvinyl alcohol dispersion of the respective NPs (Methods).

Figure 3a shows results obtained for samples containing only Au or PS particles. As expected, the Au NPs show a localized surface plasmon resonance (LSPR) around 550-560 nm, whereas the PS's scattering amplitude approximately scales with the anticipated $\lambda^{-2}$ dependence. As we move through the Au NPs' LSPR we observe a phase change of $\Delta\varphi \sim 2.4$ rad, while the PS's phase changes by approximately 0.8 rad. These results suggest that a phase measurement should allow distinguishing the two types of NPs albeit their comparable scattering amplitudes. Figure 3b further strengthens this hypothesis by comparing amplitude and phase images, acquired at two different wavelengths, for a mixed Au/PS sample. While the amplitude values are insufficient for separating Au from PS NPs, the phase-change reveals that two of the particles are resonant Au NPs. Indeed, wavelength dependent phase and amplitude measurements, based on a mixed sample with a total of 81 NPs (Figure 3c), allow recovering results reminiscent of the separate measurements (Figure 3a) by classifying the individual NPs based on their wavelength-dependent phase response.

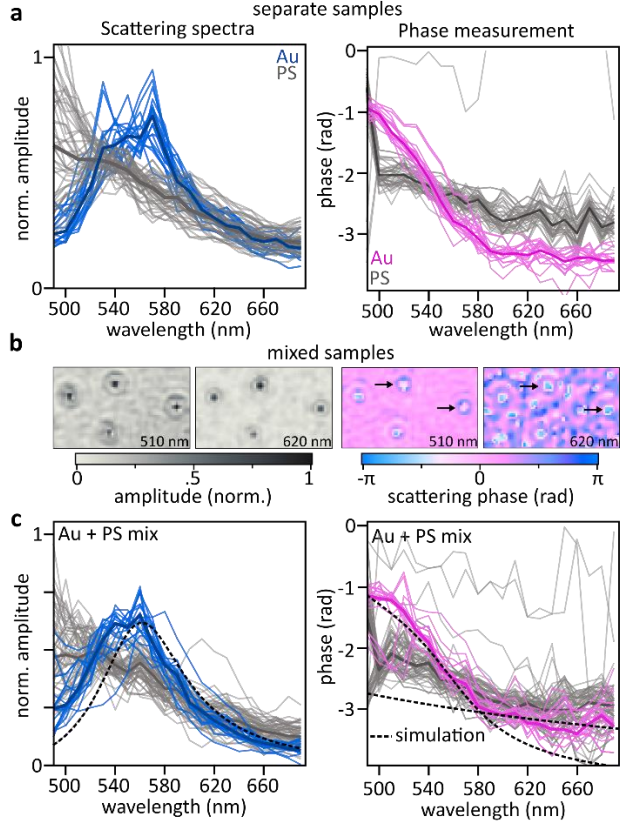

**Figure 3, Phase and scattering properties of gold and dielectric NPs**. a) Scattering amplitude (left) and phase (right) for gold (blue and pink, respectively) and PS particles (grey). The bold lines indicate median values. b) Amplitude and phase images acquired at 510 nm and 620 nm, of a mixed sample, showing two gold spheres (marked by arrows) and two polystyrene beads. c) Same as a), but for two mixed samples where PS and gold NPs are measured at the same time and are classified by their phase properties. The black dotted lines show the simulations for the spectra of the gold NPs, and the phase simulations for both gold and PS NPs.

**Additional phase shifts in reflection geometry**

The PS NP measurements presented in Figure 3 show a wavelength dependent phase that, generally, deviates from the, intuitively expected, $\varphi = \pi/2$ rad for dielectric particles due to the Gouy-phase shift. Similarly, the Au NP phase is offset and slightly deviates from the expected $\pi$ phase-change that we had previously observed for the Au nanorods using the transmission geometry.

These differences can be attributed to the fact that we are performing the measurements using a reflection geometry. Under these conditions, the center-of-mass of the individual particles is, approximately, one radius, $r$, above the glass-air interface of the sample support. As a result, the scattered wave accumulates an additional, wavelength-dependent, phase-shift with respect to the illumination wave that is reflected at the interface. We estimate this shift as:

$$\varphi_{\Delta r}(\lambda) = 2\pi \, n_{PS-PVA/Au-PVA}(\lambda) \frac{2r}{\lambda} \tag{11}$$

, where $n_{PS-PVA/Au-PVA}$ is the wavelength dependent weighted refractive index of PS or Au in the PVA matrix, assuming that the NPs' material contributes as a half-sphere in a box with dimensions $2r$ x $2r$ x $r$.

**Simulating the scattering phase**

While Equation 11 is sufficient to describe the wavelength dependent scattering phase of the dielectric PS NPs, when adding the additional $\pi/2$ Gouy-phase shift, we need to further account for the Au's LSPR to fully describe its scattering properties. We employ a simple Lorentzian resonator model to describe the LSPR, with a resonant frequency of $v_0$=535.7 THz, corresponding to $\lambda_0$= 560 nm, and a damping of $\gamma$ = 82.4 THz. The simulation parameters are chosen based on the measured scattering amplitude (Figure 3), where we describe the Lorentzian-shaped scattering amplitude as:

$$A_s(v) = \frac{1}{\pi} \frac{0.5\,\gamma}{(v-v_0)^2 + (0.5\gamma)^2} \tag{12}$$

And the scattering phase as:

$$\varphi_1(v) = atan\left(\frac{\gamma v}{v_0^2 - v^2}\right) + \pi/2 \tag{13}$$

Albeit being a simple one-frequency resonator model, the results obtained when combining Equation 11-13 near-quantitatively agree with our experimental observations (Figure 3c). In essence, the deviation from a full $\pi$ phase shift is due to the large bandwidth of the LSPR, which we only partially cover with our comparatively narrow spectral observation window. We attribute the minor discrepancies between our analytical model and the experimental data to the complex dielectric environment with the glass-polyvinyl alcohol-air interface whose exact dimensions are unknown. These variables have been shown to often considerably impact the absolute phase values[6], which complicates a quantitative description using a simple analytical model. Importantly, the directionality of the phase change observed, e.g. $\pi$ to 0 or -$\pi$ to 0, is directly related to which interference term is analyzed (Figure 1b). The two terms across the diagonal are directly related, with one being the complex conjugate of the other thus effectively inverting the directionality of the phase change.

**Spectral multiplexing: Identifying NPs beyond the phase-offset**

The dependence of the reflection geometry phase on the environment slightly hampers our ability to use the absolute scattering-phase to reliably identify metallic particles in a scattering environment, such as Au NPs inside live cells, as the environment itself impacts the measured phase. In our experience, simulating absolute scattering phases with finite difference time domain methods is highly non-trivial and even after successfully implementing such a calculation one is left with the problem that the precise scattering environment is generally unknown when measuring in a complex biological structure. Albeit these complications, the markedly different wavelength dependent quantitative phases measured for metallic and dielectric particles should, nevertheless, allow distinguishing them. Even though the absolute phase-values might shift, the relative phase-difference between two wavelengths is directly coupled to the resonance properties of the particles and, therefore, relatively unaffected by changes in the dielectric environment, albeit potential spectral LSPR changes of a few tens of nanometers in wavelength. A two-color measurement should thus allow identifying the nature of the NPs.

As our spectral-scanning approach is time-consuming and hence ill-suited for observing dynamic processes we opt for a multiplexed two-color illumination scheme where we employ two wavelengths, red- and blue-shifted with respect to the NPs' resonance around 560 nm. To implement said experiment in a single-shot manner we modify the initial imaging system (Figure 1a) by replacing the 1D diffractive beamsplitter by a 2D beamsplitter, as schematically outlined in Figure 4a. Rather than selecting a single color from our laser source we opt for collinear two-color excitation, employing wavelengths of 510 nm and 600 nm. The two-color image is detected in a colorblind fashion, as previously. To, nevertheless, distinguish the contributions of the individual wavelengths, we selectively block different colors in the diffraction orders of the beam-splitter (Figure 4a, inset). As such, we obtain single-color reference waves at distinctly different k-vectors, as seen in Figure 4a. As these reference waves interfere with the scattered electric field, they transfer the single-color image-information to distinctly different positions in momentum space (Figure 4b) which allows simple separation via Fourier filtering. Following the analysis procedure outlined above we retrieve two absolute phase values for each particle in our widefield image, from a single camera-exposure observation.

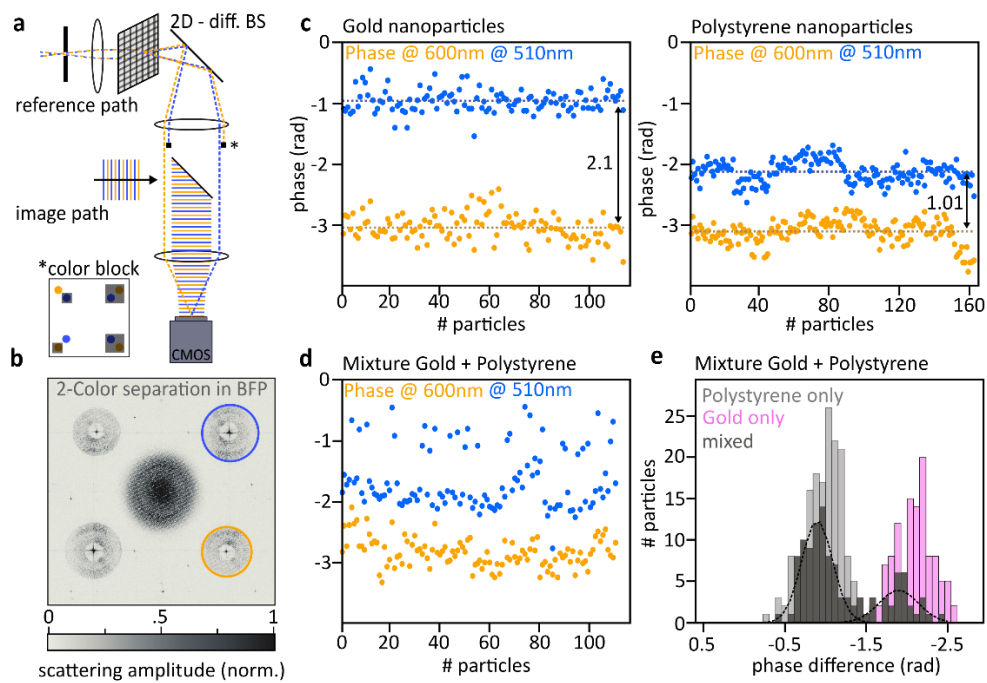

**Figure 4, One-shot detection with a 2-color interference experiment**. a) Schematic of a part of the modified setup for two-color phase detection, the two excitation wavelengths are seen in blue (510 nm) and yellow (600 nm). The reference waves are sketched as dotted lines, while the mulit-color image wave is visualized with continous lines. The diffraction orders are blocked by hard apertures, visualized as the two small black rectangles and seen in the small inset on the lower left. b) BFP image after the 2-color interference. Seen are all 4 modulated interference terms, while one component of the 510nm (600nm) interference is seen in blue (yellow). c) Measurement of gold (left) and PS (right) NPs at two differerent excitation wavelengths of 510 nm and 600 nm. d) Measurement of a mixed sample of Au and polystyrene NPs. e) Distribution of the phase difference at the two different wavlengths for each particle, plotted together with the distributions of the separate measurements from c). Two Gaussian distributions are drawn on top of the mixed histogram to visualize the two peaks.

We, initially, measure pure Au or PS samples, the results of which shown Figure 4c. For Au, based on the spectrally resolved measurements, we would have expected a phase-difference of approximately 2.4 rad (Figure 3a), which is in good agreement with the measured value of 2.1 rad. Similarly, for PS we expected 0.8 rad, and observe 1.0 rad. Finally, we measure a sample which, simultaneously, contains both PS and Au NPs and attempt distinguishing the respective particles based on our single-shot observation. The results obtained for 115 NPs are presented in Figure 4d. To better visualize these observations, we plot the phase difference at the two wavelengths, for each individual particle (Figure 4e). We observe well-separated histograms, reminiscent of a bimodal distribution, which are well described by two Gaussian distribution functions centered at approximately 0.9 rad and 1.9 rad. The relative amplitude of about 3:1 of PS (0.9 rad) to Au (1.9 rad) NPs is in good agreement with the concentrations of the respective particles prior to spin-coating.

In conclusion, we successfully implemented a widefield off-axis holographic microscope that allows determining the wavelength dependent absolute scattering phase of any nanoparticle, both plasmonic as well as dielectric. We measured the phase response of many individual particles in an intrinsically self-referenced manner by comparing the scattering-phase of interest to the, simultaneously obtained, phase of the illumination phase. Albeit being an important ingredient for performing the holographic phase measurements, the absolute phase of the reference wave is irrelevant which eliminates the need for external calibrations or complex corrections, with the only exception being the phase-term added by the semi-transparent transmission mask which is readily calculated for the relatively thick film used.

Initially, we measured in a single image the phases and scattering spectra of an array of gold nanorods of different lengths, whose phase change of approximately $\pi$ across their respective LSPR is in good agreement with the expected results. We not only observed the anticipated $\pi$-phase-change but also measure the predicted $\pi/2$ Gouy-phase contribution, for scattering off a NP, which shifts the expected $\pi$ to 0 change to - $\pi/2$ to $\pi/2$. Following these proof-of-concept measurements we determined the absolute scattering phase of small plasmonic and dielectric nanospheres and were able to readily distinguish metallic from dielectric NPs based on their wavelength dependent phase-response. To account for potential deviations in complex 3D environments and to nevertheless allow phase-based particle-identification, we devised a single shot two-color approach where the phase of every particle was simultaneously determined at two distinct wavelengths. The phase difference between these simultaneously performed experiments was then successfully used for particle-identification. We aim to use this approach in future tracking experiments in biological samples, such as gold NPs in cells, where the constant scattering environment of the specimen can be easily distinguished from the 2-color-phase properties of the NP.

## Methods

**Optical setup,** The samples of interest are mounted on a home-built inverted microscope quipped with an Olympus Apo N 60x TIRF objective (NA = 1.49) and a CMOS camera (*acA2040-90um Basler ace, Basler AG*) as detector. A spectrally filtered white-light laser (SuperK Extreme) wide-field illuminates the sample.

Reflected, or transmitted, and scattered light consecutively pass a 1:1 relay imaging system (2x AC254-200-A, Thorlabs), before being divided by a 50:50 beamsplitter to form the reference and image paths. Both paths are equipped with the same three-lens imaging system, consisting of two pairs of identical lenses in each path (AC254-150-A, AC508-250-A, Thorlabs) with a third shared lens (AC508-500-A, Thorlabs). The total magnification of the system is 100x. A semi-transmissive Fourier filter is placed into the first conjugate back-focal-plane of the image path, while the first conjugate back-focal-plane of the reference path contains a 30 μm-diameter pinhole, followed by a 2D-diffractive beamsplitter with a grating period of approx. 20 μm in the conjugate image plane[22]. An additional wedge-prism pair is placed into the image path to fine-tune the optical path-lengths of both reference and image paths.

**Transmissive Fourier Filter,** To fabricate the transmissive Fourier filter, we take a sticker the size of the one-inch glass window and laser-cut a small hole into the center. After sticking it onto the window, we evaporate a 120 nm gold layer, followed by the removal of the sticker to be left with solely a gold disc in the center of the window.

**Sample fabrication,** The nanorods used in this experiment are fabricated using e-beam lithography. A 0.17 mm cover glass is cleaned and an 8 nm-thick ITO layer evaporated on top of it, followed by a 5-min baking process at 350 degrees. Consecutively, a 180 nm-thick positive resist layer (PMMA 4%) is spin coated and baked, before exposure in the electron microscope. The developing process is performed using a 1:3 dilution of MIBK: IPA. In the final step, a 50 nm layer of gold is evaporated on top of a sample, followed by a three-hour lift-off in acetone at 50 degrees.

For the Au and PS nanoparticle samples, the cover glass (0.17 mm) is first cleaned for 10 min in acetone in the ultrasonic bath followed by 10 min in milliQ water. Consecutively the samples are exposed to oxygen plasma for about 10 min. The PS and Au nanoparticles are diluted in a PVA solution (2%), before spin coating them (1500 rpm) onto the coverslip.

**Acknowledgements**

We thank Dr. Paweł Woźniak for fruitful discussions regarding the phase accumulation in thin and thick layers. Authors acknowledge support by the Ministry of Science, Innovation, and Universities (MCIU: BES-2016-078727, RTI2018-099957-J-I00 and PGC2018-096875-B-I00) and the Ministry of Science and Innovations (MICINN: "Severo Ochoa" program for Centers of Excellence in R&D CEX2019-000910-S). N.F.v.H. acknowledges the financial support by the European Commission (ERC Advanced Grant 670949-LightNet), the Catalan AGAUR (2017SGR1369), Fundació Privada Cellex, Fundació Privada Mir-Puig, and Generalitat de Catalunya through the CERCA program.

# Supplementary material

**Gold and Palladium transmission mask measurements**

In our phase measurements of gold and polystyrene beads we accounted for the accumulated phase when passing the 120 nm gold transmission mask. To validate this phase-correction, we perform comparative measurements using either the 120 nm thick gold or an 88 nm thick palladium mask, both of which transmit approximately 0.05%. We opted for palladium as it is relatively easy to deposit and, additionally, shows a linear refractive index change over the visible/NIR spectral region, with a refractive index greater than one (Supplementary Figure 1a).

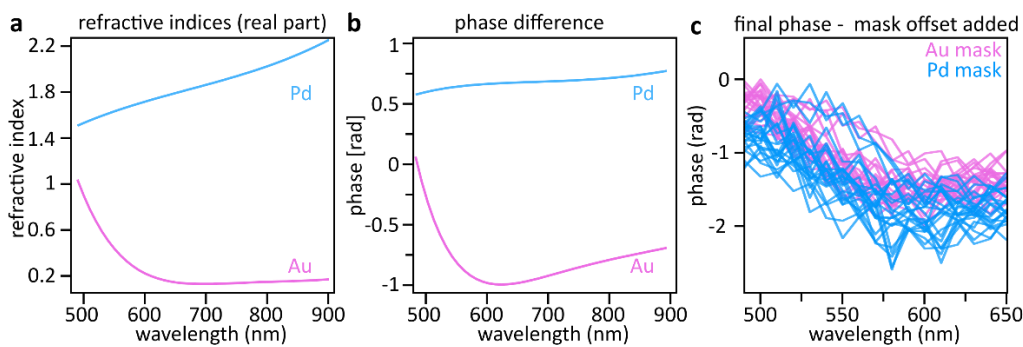

**Supplementary Figure 1, gold vs palladium mask**. a) Refractive indices of palladium (blue) and gold (pink). b) Expected phase differences acquired between a wave passing air (n=1) or the 88 nm palladium (blue) or 120 nm gold mask (pink), respectively. c) Transmission mask-corrected phase measurements performed on the same nanoparticles using the gold (pink) or the palladium mask (blue).

Supplementary Figure 1b shows the phase-difference acquired between a wave passing either air or the respective mask, calculated as:

$$\varphi_{Au/Pd}(\lambda) = 2\pi(n_{Au/Pd}(\lambda) - n_{air})\frac{d_{Au/Pd}}{\lambda}$$

The measured scattering phase has to be corrected by this wavelength dependent phase-function. To verify the correctness of this phase-correction, we perform consecutive phase measurements of the same gold nanoparticles using either the gold and palladium mask, respectively. Supplementary Figure 1c shows good agreement between the transmission mask-corrected wavelength dependent scattering phase albeit a nominal correction-difference of approximately 2 rad, thus validating our correction approach.